\documentclass{aa}

\usepackage[]{graphicx}
\usepackage[]{epsfig}
\usepackage[]{longtable}
\include{epsf}
\usepackage{natbib}
\begin{document}

\def\fdeg{\hbox{$.\mkern-4mu^\circ$}}
\def\farcs{\hbox{$.\!\!^{\prime\prime}$}}
\def\farcm{\hbox{$.\mkern-4mu^\prime$}}
\def\degr{\hbox{$^\circ$}}
\def\arcmin{\hbox{$^\prime$}}
\def\arcsec{\hbox{$^{\prime\prime}$}}
\def\sun{\hbox{$\odot$}}
\def\logg{\hbox{$\log g$}}
\def\Teff{\hbox{$T_{\rm eff}$}}
\def\fh{\hbox{$.\!\!^{\rm h}$}}
\def\fm{\hbox{$.\!\!^{\rm m}$}}
\def\fs{\hbox{$.\!\!^{\rm s}$}}
\def\hou{^{\rm h}}
\def\min{^{\rm m}}
\def\ssec{^{\rm s}}

\def\logg{log\hspace*{1mm}$g$}
\def\mfe{$\langle{\rm Fe}\rangle$}
\def\mh{$\langle{\rm H}\rangle$}

\def\unit#1{\;{\rm #1}}


\def\apj{ApJ}
\def\aap{A\&A}
\def\aaps{A\&AS}
\def\araa {RAA\&A}
\def\aj{AJ}
\def\apjl{ApJL}
\def\apjs{ApJS}
\def\asp{ASP}
\def\pasp{PASP}
\def\physrep{Phys. Rep.}
\def\apss{Ap\&SS}
\def\nat{Nature}
\def\mnras{MNRAS}


\title{The old globular cluster system of the dIrr galaxy NGC\,1427A in the Fornax 
cluster\thanks{Based
    on observations obtained at the European Southern Observatory, Chile
    (Observing Programme 70.B--0695). Partly based on archival data of the
    NASA/ESA {\it Hubble Space Telescope}, which is operated by AURA, Inc.,
    under NASA contract NAS 5--26555.}} 

\author {Iskren Y. Georgiev \inst{1,2} 
\and Michael Hilker \inst{1} 
\and Thomas H. Puzia \inst{2} 
\and Julio Chanam\'e \inst{3,2} 
\and Steffen Mieske \inst{1,4}
\and Paul Goudfrooij \inst{2}
\and Andreas Reisenegger \inst{5}
\and Leopoldo Infante \inst{5}
} 

\offprints {I.\,Y.\,Georgiev}
\mail{georgiev@stsci.edu \\
$^{**}$Founded by merging of the
Sternwarte, Radioastronomisches Institut and Institut f{\"u}r Astrophysik
und Extraterrestrische Forschung der Universit{\"a}t Bonn.}

\institute{Argelander-Institut f{\"u}r Astronomie$^{**}$, Universit{\"a}t Bonn, 
Auf dem H{\"u}gel 71, D--53121 Bonn, Germany
\and
Space Telescope Science Institute, 3700 San Martin Drive, Baltimore, Maryland 21218, USA
\and
Department of Astronomy, The Ohio State University, 140 West 18th Avenue, Columbus, OH 43210
\and
European Southern Observatory, Karl-Schwarzschild-Strasse 2, D--85748 Garching bei
M\"unchen, Germany
\and 
Departamento de Astronom\'\i a y Astrof\'\i sica, P.\,Universidad Cat\'olica de Chile,
Casilla 306, Santiago 22, Chile}

\date{Received 19 Jan 2006/Accepted 3 Feb 2006}

\titlerunning{Globular cluster system of NGC\,1427A}

\authorrunning{Georgiev, I.\,Y. et al.}

\abstract{We present a study of the old globular cluster (GC) population of the 
dwarf irregular galaxy NGC\,1427A using multi-wavelength VLT observations in 
$U,B,V,I,H_{\alpha},J,H,$ and $K_{s}$ bands under excellent observing conditions. 
We applied color and size selection criteria to select old GC candidates and made 
use of archival ACS images taken with the {\it Hubble Space Telescope} to reject 
contaminating background sources and blended objects from the GC candidates' list. 
The $H_\alpha$ observations were used to check for contamination due to compact, 
highly reddened young star clusters whose colors and sizes could mimic those of 
old GCs. After accounting for contamination we obtain a total number of 
$38\pm8$ GC candidates with colors consistent with an old ($\sim$\,10 Gyr) and metal-poor 
($Z<0.4\times Z_{\odot}$) population as judged by simple stellar population models. 
Our contamination analysis indicates that the density distribution of GCs in the 
outskirts of the Fornax central cD galaxy NGC\,1399 may not be spherically symmetric. 
We derive a present-day specific frequency $S_N$ of 1.6 for NGC\,1427A, a value 
significantly larger than what is observed in the Local Group dwarf irregular 
galaxies and comparable with the values found for the same galaxy types in the 
Virgo and Fornax clusters. Assuming a universal globular cluster luminosity function  
turnover magnitude, we derive a distance modulus to NGC\,1427A of $31.01\pm0.21$\,mag 
which places it $\sim3.2\pm2.5\,({\rm statistic})\pm1.6\,({\rm systematic})$\,Mpc 
in front of the Fornax central cD galaxy NGC\,1399. The implications of this result 
for the relationship between NGC\,1427A and the cluster environment are briefly discussed.
}

\maketitle



\section{Introduction}\label{intro}

Old Globular Clusters (GCs) are complexes of stars which have been formed almost 
simultaneously out of material with basically the same initial chemical composition. 
They form during major star formation episodes in virtually every galaxy.Hence 
they are regarded as a fossil record of the initial conditions of the early host 
galaxy formation history.

With respect to the formation and/or assembly of early-type galaxies and their
globular cluster systems (GCSs) three main competing scenarios emerged. The first 
assumes a hierarchical build-up of massive galaxies from a (large) population of 
pregalactic gaseous cloud fragments. In this scenario two major epochs of star 
formation occurred, with both the blue (metal-poor) and the red (metal-rich) GC 
populations formed {\it in situ} with a pause in between the two bursts of GC 
formation \cite[]{Forbes97}. 
Another possibility,within the same scenario, is to have a long series of smaller 
star-forming events with the metal-poor clusters as first generation and the metal-rich 
ones formed in mini-mergers at high redshifts \cite[]{Beasley02}. The second scenario 
regards major mergers of disk galaxies \cite[]{Schweizer87, Ashman&Zepf92} in which 
the metal-poor GC population form early in ``Searle-Zinn'' fragments in the halos of 
the progenitor disk galaxies, while most of the metal-rich GCs form during later 
progenitor mergers, and thus are younger. The third scenario, dissipationless satellite 
accretion, assumes that the host galaxy forms by classical monolithic collapse and 
then accretes smaller galaxies \cite[]{Cote98}. The GCs of these dwarf galaxies are 
being captured and/or tidally stripped \cite[]{Hilker99, Cote02}. For a thorough 
discussion of the GCS formation scenarios we refer the reader to the reviews by 
\cite{Ashman&Zepf98}, \cite{Carney&Harris98}, \cite{Elmegreen99}, \cite{Kissler-Patig00}, 
\cite{vandenBergh00} and \cite{Harris03}. 

In all these models, the GCSs of low-mass dwarf galaxies, the most numerous galaxy 
type in galaxy clusters \cite[][and references therein]{Sandage05}, are envisioned 
as the building blocks of the GCSs of the more massive galaxies. At present the role 
of the GCSs of dIrr galaxies in the frame of galaxy formation is unclear. A recent 
study by \cite{Sharina05} compared the GCS properties of dE, dSph, and dIrr in field 
and group environments and found that all three dwarf galaxy types host a population 
of old GCs with very similar $(V-I)$ colors. The mean color of this population is almost 
identical to the $(V-I)$ colors of the blue GC population in massive early-type galaxies. 
One way to address the evolutionary connection between dE, dSph, dIrr, and massive 
galaxies is to study the specific frequency $S_N$ \cite[the number of GCs ($N_{GC}$) 
normalized by the galaxy's luminosity,][$S_{N}=N_{GC}10^{0.4(M_{V}+15)}$]{Harris&vandenBergh81} 
of such galaxies. Generally, the mean $S_{N}$ values of dE, dSph, and dIrr galaxies 
in the Virgo and Fornax cluster are comparable \cite[]{Seth04} and very similar to 
$S_N$ values of giant elliptical galaxies, which implies that all form GCs in the same 
proportion (per galaxy magnitude). However, the $S_{N}$ of dwarf galaxies in clusters 
is significantly higher than for dE, dSph, and dIrr in the field environment \cite[eg. 
$S_{N}=0.5$ for LMC,][]{Harris91}. The mechanisms that drive the transition between 
the GCSs of field and cluster dwarf galaxies are unclear.

Here we study the globular cluster systhem of NGC\,1427A, is the brightest dwarf irregular 
galaxy in the Fornax galaxy cluster. Throughout this work we adopt a distance modulus 
of $(m-M)=31.39\pm0.20$\,mag ($19\pm1.8$\,Mpc) to the Fornax cluster, as measured with 
Cepheids \cite[HST Key Project,][]{Freedman01}. NGC\,1427A has a mean radial velocity 
of $2027.8\pm0.8$\,km/s \cite[]{Bureau96}, which is $\sim600$\,km/s higher than that of the 
cD elliptical NGC\,1399 ($1430\pm9$\,km/s), located at the cluster center, and twice 
larger than the cluster velocity dispersion \cite[$\sigma_{v}=325$\,km/s,][]{Ferguson90}. 
Such a large peculiar velocity is shared also by other cluster members, such as NGC\,1404, 
as shown by the double-peaked redshift distribution of the Fornax cluster and the distinct 
group of dwarf galaxies falling toward the cluster center \cite[]{Drinkwater01}. NGC\,1427A 
shows a ring-like pattern of star formation with prominent distinct starburst complexes 
ionizing their immediate surroundings. One of the first studies of NGC\,1427A by \cite{Hilker97}, 
relying on its morphological appearance and the apparent proximity to the cluster center, 
suggested that the interaction with the cluster environment is responsible for the observed 
active star formation and its future morphological evolution. \cite{Cellone97} envisioned 
a collision with one of the many dwarf ellipticals populating the cluster center to be 
responsible for the ring-like appearance of NGC\,1427A and speculated that the bright complex 
in the Northern part of the galaxy could be the intruder. A study by \cite{Chaname00} of 
the ionized gas kinematics of NGC\,1427A showed that this Northern object shares the same 
general kinematical pattern as the rest of the galaxy body, hence making unlikely its external 
origin. They also argued that the most likely scenario explaning the morphological features 
of NGC\,1427A is due to its passage through the hot intracluster medium of Fornax. Overall, 
this galaxy represents an ideal target to probe the influence of the cluster environment 
on its globular cluster population. In order to more firmly study the role of the dwarf 
galaxies in the context of galaxy and GCS formation scenarios, and the environment impact 
on their GCSs, a larger sample of dwarf galaxies should be considered, which is our goal for 
a future study.

In Section\,\ref{obsred} we describe the observational data and its reduction.
Section\,\ref{gccs} is devoted to the globular cluster candidate selection 
and contamination estimates, and in Sect.\,\ref{analysis} we analyse the properties 
of the selected GC candidates. Finally, in Sect.\,\ref{conclusions} the main 
conclusions are presented.

\section{Observations, reduction and photometry}\label{obsred}

\subsection{Observational data}\label{selobs}

Deep optical images of NGC\,1427A in $U,\ B,\ V,\ I$, and H$\alpha$ band passes were 
taken with the Focal Reducer and Low Dispersion Spectrograph 1 (FORS1) on the UT2 of 
the ESO Very Large Telescope (VLT) at Cerro Paranal, Chile. In addition, near-infrared 
(NIR) images in $J$-band (1.5\,$\mu$m) were taken with the Infrared Spectrometer And 
Array Camera (ISAAC) attached to the UT1 of the ESO/VLT. The complete NIR data set 
includes $H$ and $K_s$ imaging, however only the $J$-band was deep enough to perform 
this study. The FORS1 instrument has a $2048\times2048$ CCD detector with 0\farcs2/pixel 
resolution providing a 6\farcm8\,$\times$\,6\farcm8 field of view. The ISAAC imager 
is equipped with $1024\times1024$ array with pixel scale of 0\farcs15/pixel covering 
a 2\farcm5\,$\times$\,2\farcm5 field. Two adjacent fields were taken in the NIR in 
order to cover NGC\,1427A. All observations of NGC\,1427A were performed under photometric 
conditions.
\begin{table*}[ht]
\centering
\caption{\label{log}Log of observations.}
\begin{tabular}[l]{clcccc}
\hline\hline
Date	   &	Filter	 &	Center Position     &	Exp.time	& FWHM& Seeing\\
	   &		 &	RA; DEC (J2000)	    &	[sec]		&[pixels]& [\arcsec]\\
\hline
2003-09-29 &	U	 & 03:40:10.0\ \ -35:36:57.9& $7  \times 1595$	& 3.0 & 0.60 \\
2002-11-06 &	B	 & 03:40:10.0\ \ -35:36:57.9& $8  \times 450$	& 2.4 & 0.48 \\
2002-11-02 &	V	 & 03:40:10.0\ \ -35:36:57.9& $10 \times 150$	& 2.9 & 0.58 \\
2002-11-02 &	I	 & 03:40:10.0\ \ -35:36:57.9& $20 \times 165$	& 2.7 & 0.54 \\
2003-01-02 &H${\alpha}$  & 03:40:10.0\ \ -35:36:57.9& $6  \times 1000$	& 2.7 & 0.54 \\
2002-10-05 &	J	 & 03:40:15.0\ \ -35:37:14.0& $30 \times 40$	& 2.8 & 0.42 \\
2002-11-06 &	J	 & 03:40:05.0\ \ -35:37:14.0& $24 \times 40$	& 3.3 & 0.49 \\
\hline
\end{tabular}
\end{table*}

The observation dates, the exposure times and the average seeing are listed 
in Table\,\ref{log}. The seeing (FWHM) was measured on the combined final images using 
the IRAF\footnote{IRAF is distributed by the National Optical Astronomy Observatories, 
which are operated by the Association of Universities for Research in Astronomy, Inc., 
under cooperative agreement with the National Science Foundation.} routine PSFMEASURE. 
The different seeing in the Ks band for the two fields (observed in the same night) is 
mainly due to the low number of stars used for its computation. 

The six H$\alpha$ images were obtained with the FORS1 H$\alpha$/2500+60 filter with 
central wavelength at $\lambda=6604$\,{\AA} and FWHM\,$=64$\,{\AA}.
Given the radial heliocentric velocity of $V_{r}=2027.8\pm0.8$\,km/s \cite[]{Bureau96}, 
NGC\,1427A is placed at redshift $z=0.00676$. Hence, the H$\alpha$ emission lines 
of NGC\,1427A (rest frame wavelength at 6563\,{\AA}) are redshifted by 44\,\AA, 
which is 3\,\AA\ above the central wavelength of the H$\alpha$ filter used.

\subsection{Reduction and photometry}\label{redphot}

The basic image reduction was performed in a standard way using IRAF software 
packages. The bias subtraction and flat field corrections were applied using the CCDRED 
package taking into account the CCD noise parameters. The IR images were taken in the 
usual object-sky-object-sky sequences. Both object and sky exposures were bias- and 
flatfield-corrected in the same manner. The sky images were cleaned from all detected 
sources with a SExtractor object map, and object residuals were manually replaced with 
nearby empty sky patches. Then, these cleaned sky images were subtracted from the 
associated object exposures to obtain a background corrected science frame. Before 
combination, each single exposure in each filter was shifted by integer pixels to a 
defined reference image to provide the image flux conservation during the transformations. 
For cosmic rays rejection we used the laplacian cosmic-ray identification algorithm 
described and provided as IRAF cl-script\footnote{http://www.astro.yale.edu/dokkum/lacosmic}  
by \cite{vanDokkum01}. 
\begin{figure*}
\begin{center}
\epsfig{figure=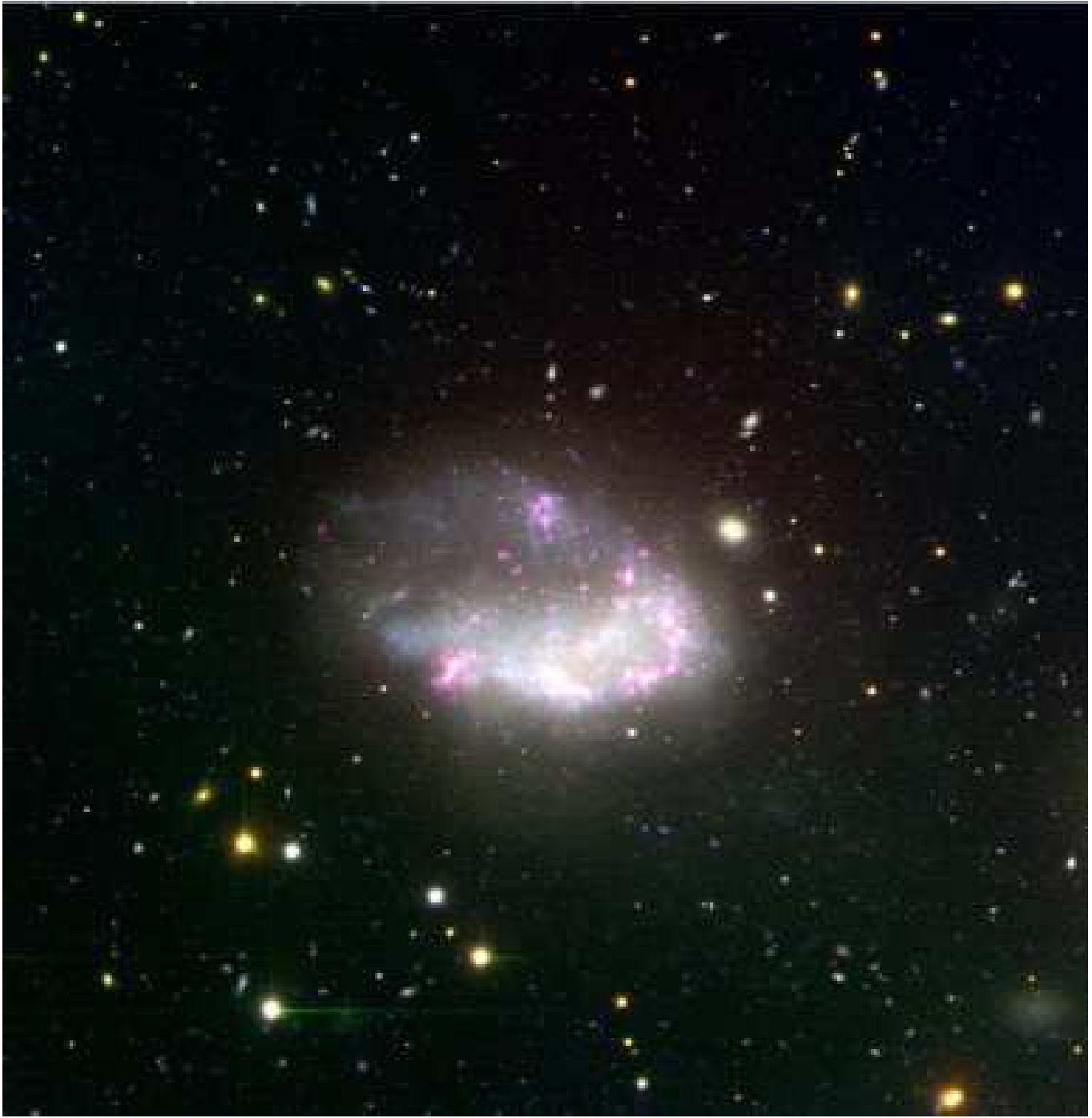, width=180mm}
\caption{\label{color}NGC\,1427A color composite image. For the blue, green and red 
channels the $U$, $V$ and H$\alpha$ filters were used, respectively. North is up, 
East is on the left. The entire field of view of this image is
$\sim6\farcm7\times6\farcm7$  
which corresponds to $\sim37\times37$\,kpc at the Fornax Cluster distance.}
\end{center}
\end{figure*}
Examination of the output rejected pixel map showed that this routine did a nice 
work cleaning all cosmic rays without touching the sharp tips of the point sources. 
Given the sharpness of the images (see FWHM in Table 1) this was a major problem 
when using the IRAF built-in rejection algorithms, which cut the stars' tips, 
during the image combining process. This is why we preferred first to clean the 
reduced images in each filter and then to combine them to a single image without 
rejection algorithms. From the final average combined images a color image of 
NGC\,1427A was produced (Fig.\,\ref{color}) using $U$, $V$ and H$\alpha$ filters.

Point sources in NGC\,1427A were identified using the DAOFIND routine of the 
DAOPHOT package \cite[]{Stetson87}, adopting a detection threshold of 5\,$\sigma$ 
above the sky level. Lower thresholds were found to give too many spurious detections. 
Visual image inspection showed that no objects were missed and only few false detections 
were included in the lists.

In order to improve the faint object photometry the underlying and extended 
galaxy light was subtracted form the original images. This was done with a 
ring aperture median filtering on images from which bright objects had been
previously subtracted. A smoothing kernel of 41 pixels radius proved to give 
the best results, i.e., the wings of the bright, well-exposed stars were not 
affected. This was confirmed by the comparison between magnitudes of stars 
measured in the subtracted and non-subtracted images (see below). The filtered 
images in each pass-band were subtracted from the original ones, and the 
resulting images then were used for photometry. Due to the complex structure 
of this irregular galaxy it was impossible to fit and subtract a simple model.
\begin{figure*}[ht]
\begin{center}
\epsfig{figure=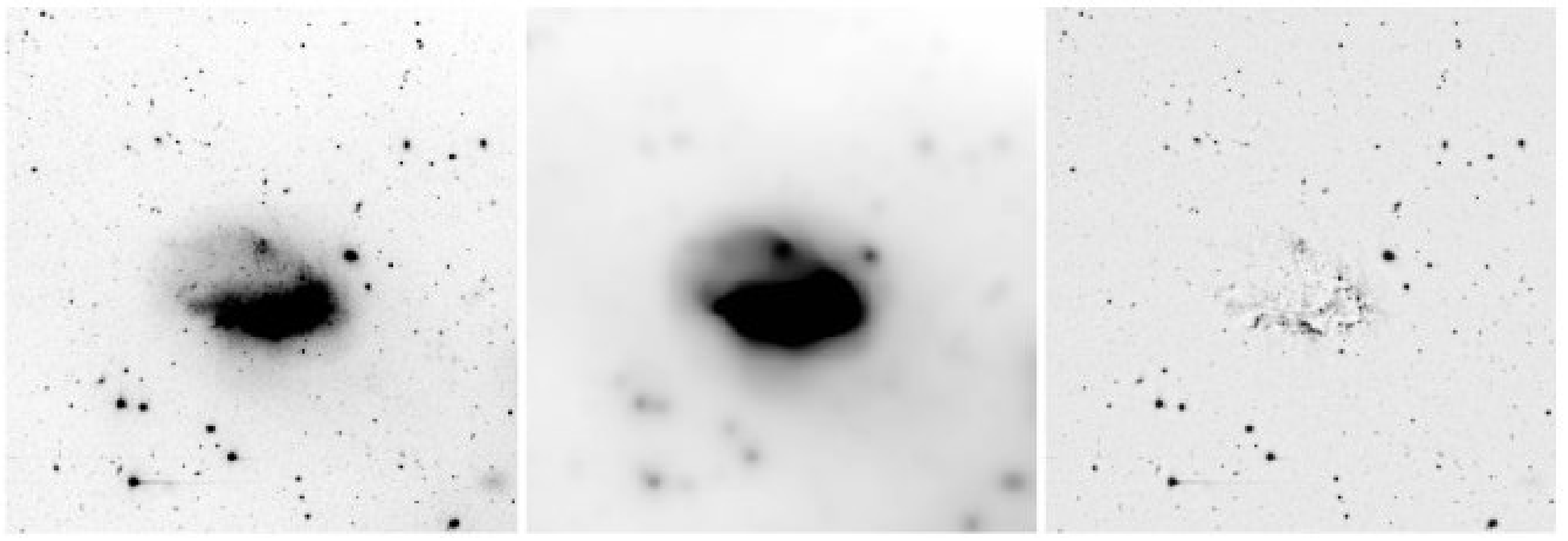, width=62mm, angle=270}
\caption{\label{smooth}Left: $V$ band image of NGC\,1427A. Middle: median-filter 
smoothed image (using smoothing kernel of 41 pixels radius). Right: median-filter 
subtracted residual image.} 
\end{center}
\end{figure*}

An example of the V-band images passed through the filtering procedure is shown in 
Fig.\,\ref{smooth}. With this choice of ring radius only the brightest overexposed 
foreground stars and extended background galaxies were affected. The residual image 
on the right panel was used for the final photometry. In order to obtain reliable 
photometry for all objects, especially those in the most crowded regions, a point-spread 
function (PSF) photometry was performed. To create the PSF model we selected between 
7 and 15 (depending on the filter) well-isolated PSF stars. The PSF photometry was 
performed with the DAOPHOT/ALLSTAR procedure.

To account for the star light lost due to the PSF model fitting and smoothed image 
subtraction, the differences between the aperture photometry 
and the PSF photometry magnitudes were determined using curves of growth for various 
aperture sizes in all filters. These curves of growth proved that the median filtering, 
with this choice of of kernel raduis, did not affect the stellar profiles significantly. 
Hence, the correction to be applied is mainly due to PSF-model fitting effects. 
We adopted a radius of 16 pixels (corresponding to $>5\times$FWHM ) for all filters. 
The corrections are listed in Table\,\ref{psfcor}.

\begin{table*}[ht]
\centering
\caption{\label{psfcor}Listed are the PSF magnitude corrections ($r_{ap}=16\rightarrow\infty$); 
the accuracy of the fits to the transformation equations and the completeness limits in 
$U,B,V,I$ and $J$ bands.}
\begin{tabular}[l]{ccccccc}
\hline\hline
Filter	 	& $U$ &	$B$ & $V$ & $I$ & \multicolumn{2}{c} {$J$}\\
		&     &	    &     &     & Field1	& Field2	\\
\hline
Corr.		& $-0.099$ & $-0.058$ & $-0.087$ & $-0.074$ & $-0.074$ & $-0.057$ \\
rms		&  $0.058$ &  $0.042$ &  $0.023$ &  $0.026$ &  $0.050$ &  $0.045$ \\
90\%		&  $26.21$ &  $26.95$ &	 $25.65$ &  $25.66$ &  $21.12$ &  $20.67$ \\
50\%		&  $26.85$ &  $27.41$ &	 $26.55$ &  $26.28$ &  $21.99$ &  $21.59$ \\
\hline
\end{tabular}
\end{table*}

Photometric calibrations of each data set from instrumental to standard magnitudes 
were performed in a standard way using \cite{Landolt92} standard stars which were 
observed in the corresponding nights. After fitting the zero points, extinction 
coefficients and color terms (as given by the ESO Quality Control for the corresponding 
nights), the resultant rms of the fits for the $UBVI$ and $J$ photometry are shown 
in Table\,\ref{psfcor}.

The main objects of interest for the current paper are the old globular clusters 
associated with NGC\,1427A. At the adopted distance of the Fornax Cluster, 
one FORS1 pixel (0\farcs2) corresponds to $\sim19$\,pc projected size. Given the mean FWHM 
($\leq3$\,pixels) of our images all compact sources associated with NGC\,1427A 
(SCs, rich OB associations, compact H\,{\sc ii} regions, or very luminous giant 
stars) are expected to have a star-like profile and the adopted PSF photometry 
is a good choice.

\subsection{Completeness determination}\label{compl}
At the distance of the Fornax Cluster the expected universal turnover magnitude 
of the globular cluster luminosity function \cite[absolute value $M_{V,TO}\simeq-7.4$\,mag;][]{Harris96, Harris01} 
is expected at $V_{TO}\sim24$ mag. To estimate up to which limiting magnitude our observations are 
complete we used the final combined and background subtracted images and the 
IRAF/DAOPHOT/ADDSTAR procedure.

\begin{figure}[ht]
\begin{center}
\epsfig{figure=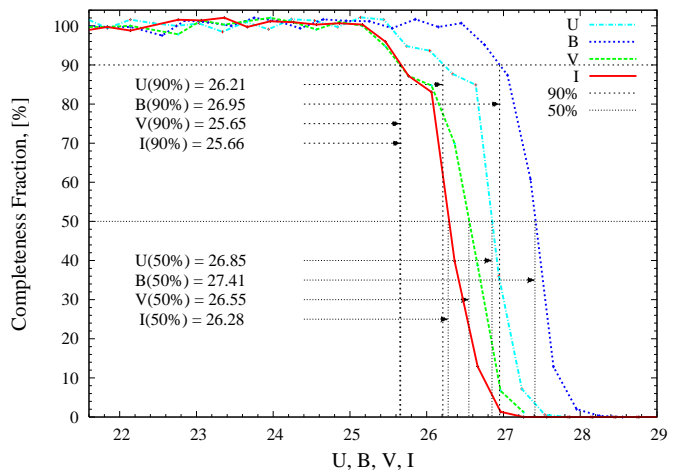, width=91mm}
\caption{\label{completeness}Completeness functions estimated with the add-star 
experiment.}
\end{center}
\end{figure}

To avoid artificial crowding in the images we added 100 stars per image with 
magnitudes ranging from 21 to 29 mag in the optical and 14 to 25 mag in the IR 
images in 100 runs. Thus, the total number of stars that were added per filter is 
10\,000. The artificial objects were generated from the PSF of each correspondent 
field. Detection and photometry of the artificial objects was performed in the 
same manner as described in Sect.\,\ref{redphot}. The recovered number of objects 
divided by the number of the input artificial stars then gave us the completeness 
fraction. In Fig.\,\ref{completeness} we show the completeness functions for the 
optical and IR data. The completeness is a function of the background level, hence 
it is expected to be variable toward the central galaxy regions. This effect should 
be accounted for if the photometric study is devoted mainly to the objects in those 
inner most regions. However, the GCs which we study here are typically located at 
large galactocentric distances and this effect is expected to be not so significant. 
The long exposure times and the photometric conditions resulted in deep images in the 
optical wavelengths (see Fig.\,\ref{completeness} and Table.\,\ref{psfcor}). In the 
$UBV$ and $I$-bands we easily reach $>1$\,mag beyond the expected turnover of the GC 
luminosity function , while in the near-IR we reach only its bright end 
\cite[e.g., $M_{J,TO}=-9.21$\,mag;][thus $J_{TO}\simeq22.2$\,mag]{Barmby01}.

\subsection{H$_{\alpha}$ calibration and results}\label{Hacalib}
The calibration of the H$\alpha$ images was based on the spectro-photometric standard 
star LTT\,1020 \cite[]{Hamuy92,Hamuy94}. As initial guess for the extinction coefficient 
we used the observatory value for the $R$ filter and then computed our zero point. 
Using the magnitude and luminosity of LTT1020 we obtained the following relation
\begin{equation}
m_{\rm H_{\alpha}}=-2.5\ {\times}\ \rm log(F_{\rm H\alpha})-21.51\pm0.11
\end{equation}
where F$_{\rm H\alpha}$ is the flux in erg s$^{-1}$\,cm$^{-2}$\,\AA$^{-1}$. 
This relation was subsequently used to convert from magnitude to H$_{\alpha}$ 
luminosity in order to derive the star formation rate (SFR) for NGC\,1427A (see below).

The raw H$\alpha$ images were reduced, aligned, registered, and a final combined 
and background subtracted H$\alpha$ image was produced (Sect.\,\ref{redphot}). 
In order to subtract the continuum emission from our H$\alpha$ image we follow  
\cite{Knapen04} and use as reference the $I$-band image (due to the lack of 
a $R$-band image). To determine the scaling factor, which tells how the continuum 
image must be scaled to match the intensity level of the continuum emission in 
the H$\alpha$ image, we followed the approach described by \cite{Boker99}.

We applied the so determined scaling relation to the $I$-band image and subtracted 
it from the H$\alpha$ image. This gave us the resulting pure H$\alpha$ emission-line 
image. As expected, sources without H$\alpha$ emission disappeared to the background  
level whereas sources having H$\alpha$ in absorption were over-subtracted. A check 
for young embedded star clusters or compact H\,{\sc ii} regions contaminating the 
GC candidate sample revealed no emission in excess or in absence at the GC candidate's 
positions on the residual H$_{\alpha}$ images. Our conclusion is that the selected 
GC candidates are hardly if at all polluted by young compact sources from the inner 
starburst galaxy regions.

The brightest background/foreground objects were masked out on the so obtained 
residual H$\alpha$ images and the H$\alpha$ magnitude of the galaxy was measured 
using the IRAF/STSDAS task ELLIPSE. For the conversion from magnitude to luminosity 
we used the adopted distance of 19\,Mpc \cite[]{Freedman01}. The star formation 
rate was derived from the H$\alpha$ flux using the improved relation given by 
\cite{Hunter&Elmegreen04}: 
\begin{equation}
\dot{M}=5.96\times10^{-42}L_{H_{\alpha}}10^{0.4A_{H_{\alpha}}}\ {\rm M}_{\odot}/{\rm yr}
\end{equation}
where $L_{H_{\alpha}}$ is the $H_{\alpha}$ flux in erg/s, $A_{H_{\alpha}}=0.811A_{V}$ 
is the extinction coefficient in $H_{\alpha}$. We derived $\dot{M}=0.057$ M$_{\odot}$/yr. 
This is a typical value for a starburst dIrr galaxy \cite[]{Hunter&Elmegreen04, Seth04}. 
This estimate should be regarded as an upper limit for the SFR since, due to the lack 
of flux calibrated emission-line spectra for N\,1427A, no care was taken for the N\,II\,6548,6583 
emission-line doublet, which partly falls within the $H_{\alpha}$ filter pass-band.

\subsection{Foreground and intrinsic extinction}\label{fge}
Absorption caused by the Galactic interstellar medium (ISM) that affects the objects' 
magnitudes, hence colors, have to be taken into account in order to compare the 
object's colors (magnitudes) with those of stellar evolutionary models (Sect.\,\ref{cms}).

In order to calculate the extinction correction towards NGC\,1427A for each filter 
we adopted\footnote{\label{ned}http://nedwww.ipac.caltech.edu/} $E(B-V)=0.012$ 
\cite[]{Schlegel98}, $A_{V}=3.1 \times E(B-V)$, formulae (2) and (3) in 
\cite{Cardelli89} and the central wavelengths for the FORS1 and ISAAC filters taken 
from the ESO web page.

Although the purpose of the $J,H,K$ imaging was to determine the extinction map of 
NGC\,1427A, the images turned out to be too shallow for this goal. Therefore, we can not 
apply individual internal reddening correction, which could be large for clusters 
located inside or behind the galaxy body. Nevertheless, we do not expect this to be a 
significant effect because of the following reasons. First, as will be seen in Fig.\,\ref{pos}, 
we observe $\sim10-15$ candidates in the innermost galaxy regions, and only a fraction 
of these are expected to be inside or behind, and therefore affected by internal extinction. Second, 
\cite{James05}, studying the extinction on the H$_{\alpha}$ flux as a function of the 
galaxy type, show that the mean extinction correction for dIrrs is $A_{\rm H_{\alpha}}\simeq0.4$\,mag, 
which transforms to $E_{B-V}\simeq0.06$ (or $V-I\simeq0.07$\,mag). The overall internal 
reddening therefore is expected to be small and this is why we consider in the following 
only the foreground extinction.

\section{Globular cluster candidates selection}\label{gccs}
Our final photometry list contains stars, star clusters and extended objects (galaxies, 
H\,{\sc ii} regions, OB-associations, etc.). In order to select globular cluster 
candidates we apply color, magnitude and size selection criteria according 
to the expected properties of these old stellar systems at the distance of NGC\,1427A.

\subsection{Color and magnitude selection}\label{cms}
The globular cluster candidate selection is mainly based on the colors of the objects 
in the color-color diagrams. The adopted distance modulus to the Fornax cluster
limits the expected brightest globular cluster magnitude to be $m_{V}\simeq21$\,mag 
assuming $M_{V}=-10.29$ for the brightest Milky Way (MW) globular cluster $\omega$\,Centauri 
\cite[]{Harris96}.
\begin{figure*}[ht]
\epsfig{figure=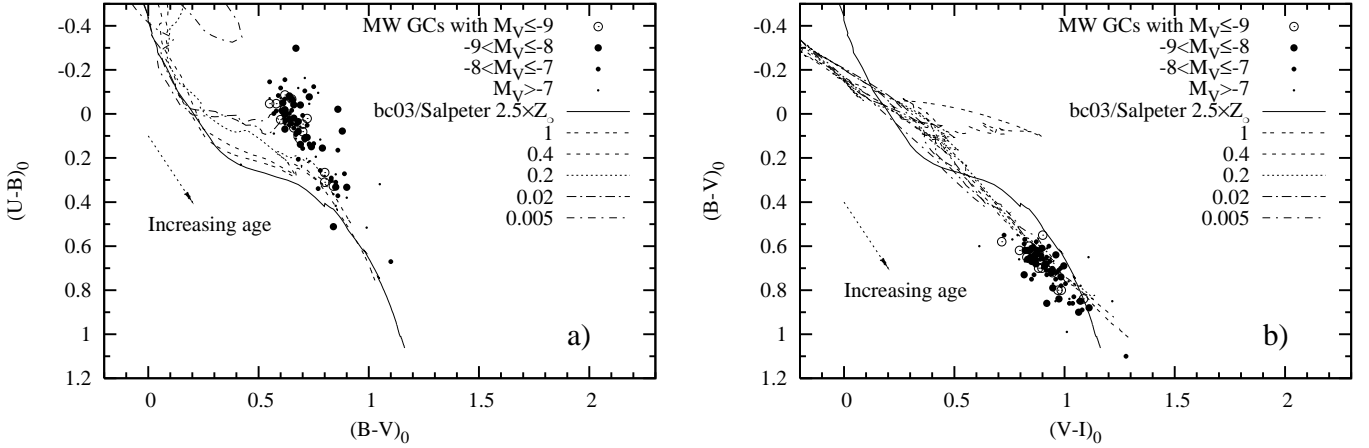, width=185mm}
\caption{\label{mwgc2}Milky Way globular clusters' $(U-B)_0$\,vs.\,$(B-V)_0$ 
({\bf a}) and $(B-V)_0$\,vs.\,$(V-I)_0$ ({\bf b}) color-color 
the power of a color-color selection''. (readers can see that anyway) diagrams. The 
clear metallicity separation of the MW globular cluster system can easily be seen in 
({\bf a}), while in ({\bf b}) the objects' metallicity separation smears out and the 
SSP models for different metallicities are indistinguishable.}
\end{figure*}

\begin{figure}[ht]
\begin{center}
\epsfig{figure=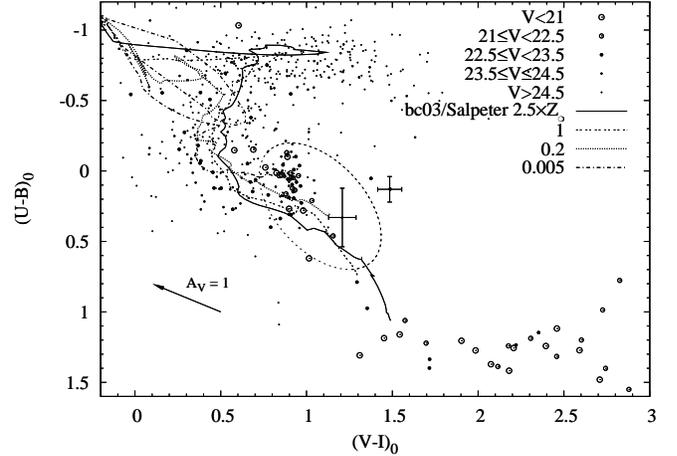, width=90mm}
\caption{\label{full}$(U-B)_0$\,vs.\,$(V-I)_0$ diagram showing the colors of all 
point sources (Sect.\,\ref{pssc}) found in the FORS1 field. The different point types 
indicate objects with different magnitudes. The arrow shows the magnitude and the 
direction that would have to be applied to the colors of objects subject to an intrinsic 
extinction of $A_{V}=1$\,mag. With lines are plotted the \cite{BC03} SSP models for 
different metallicities. Representative photometric errors are shown for two GCs 
with luminosities around the expected GCLF turnover magnitude.}
\end{center}
\end{figure}

To constrain a region of typical GC colors we use the 2003 updated version of the 
\cite{Harris96} catalog of MW GCs \footnote{http://www.physics.mcmaster.ca/Globular.html}. 
The color selection limits, which we chose as first selection criterion, are defined as an ellipse 
that approximately encompasses the colors within $-0.16\leq(U-B)\leq0.7$ and 
$0.72\leq(V-I)\leq1.4$ (see Fig.\,\ref{full}). These color ranges are found to be typical 
for GCs in a large variety of galaxy types \cite[eg.][and ref's therein]{Ashman&Zepf98, 
Larsen01, Lotz04, Chandar04, Sharina05}. 

The $(U-B)$\,vs.\,$(V-I)$ and the $(U-B)$\,vs.\,$(B-V)$ color combinations proved to be 
the most distinctive when disentangling ages and metallicities. Any color combination 
{\it not} including the $U$-band makes the models almost indistinguishable (compare the 
two panels in Fig.\,\ref{mwgc2}), hence leading to large uncertainties in assessing ages 
and/or metallicities. Due to line blanketing effects in the stellar atmospheres the 
$(U-B)$ color is very sensitive to metallicity variations. Therefore, inclusion of the 
$U$-band in photometric studies aiming at accurate age/metallicity determinations is very 
important. A detailed discussion on the right filter selection when comparing observations 
with evolutionary synthesis models is provided by \cite{Anders04}. Indeed, as shown in 
Fig.\ref{mwgc2}, the well established bimodal metallicity distribution of the MW globular 
cluster system \cite[e.g.,][and references therein]{vandenBergh93, Carney&Harris98, Mackey05} 
can easily be seen when using $(U-B)$ color.

In Fig.\,\ref{full} we show the colors of all 
point sources detected in the FORS1 field, corrected only for foreground Galactic 
extinction. The different symbol types indicate different $V$-band magnitudes of the
objects. The arrow shows the direction and the magnitude with which the color of an 
object would have to be shifted if an extinction of $A_{V}=1$\,mag intrinsic to 
NGC\,1427A is applied.

Representative photometric errors are shown in Fig.\,\ref{full} to demonstrate 
the accuracy of our photometry for objects in the magnitude bins around the expected 
GC luminosity function turnover at $\sim24$\,mag and obeying the GC candidate selection 
criteria (see Sect.\,\ref{pssc}).

In the color-color diagram one can see a large number of blue objects. Examination 
of the positions of these objects showed that most of them are found in the inner 
starburst regions of NGC\,1427A. The rest of the blue objects are associated with 
faint sources, likely background galaxies and sources associated with them: An 
example is the background spiral galaxy westward of NGC\,1427A (Fig.\,\ref{color}). 
The visual inspection of the observed very red and bright objects with $(V-I)>1.5$ 
showed that they are saturated foreground stars and likely background elliptical 
galaxies. In the color region where GCs are expected a high density of objects is 
observed.

\subsection{Point source selection criterion}\label{pssc}
As pointed out in Sect.\,\ref{redphot} the distance to NGC\,1427A puts an additional 
constraint to the expected sizes of the GCs associated with NGC\,1427A. The typical 
GC half-light radii are in the range 1 to 20\,pc \cite[eg.][]{Kundu01, vandenBergh04, 
Jordan05}. Given the adopted distance to the Fornax cluster, 1\,pixel of our images 
corresponds to $\sim19$\,pc projected size. Therefore, GCs are expected to be 
unresolved and to have point source radial profiles. As point source selection 
criterion we used the IRAF ALLSTAR task output parameter ``sharp''. Sharp is 
computed during the fit of the PSF model and estimates the intrinsic angular size 
of the object. It is roughly defined as the difference between the square of the 
width of the object and the square of the width of the PSF. It has values close 
to zero for unresolved point sources, large positive values for blended sources and 
partially resolved galaxies and large negative values for cosmic rays and blemishes. 

As first selection criterion we used the color-color limits explained in Sect.\,\ref{cms}. 
Then for the color-color selected objects we plotted the $V$-band sharp value against 
$V_0,\ (U-B)_0$ and $(V-I)_0$. Plots of the sharp values for the other filters were 
produced and showed insignificant differences. However, such differences might be 
expected between $U$-band and $V$-band sharp values for example if in the selected 
color-color region highly reddened bright compact H\,{\sc ii} regions embedded in the 
NGC\,1427A starburst regions are observed. Such objects can have different profiles 
in the $U$ and in the $V$ filter due to the enhanced contribution of the ionized gas 
in $U$-band which is invisible in $V$.

The shape of the functions defining the upper and lower cutoff sharp values were 
determined from the sharp value distribution of artificially added stars previously 
used for the completeness tests (Sect.\,\ref{compl}). The scatter of the sharp value 
increases towards fainter magnitudes due to increasing uncertainties in the PSF 
fitting procedure. At approximately $V_0=24.3$ and $(V-I)_0=0.86$ mag couple of objects 
slightly above the sharp cut-off function are observed. A visual inspection of these 
objects revealed that they are somewhat extended sources embedded in the central 
starburst regions, which suggests at their young nature. An additional check was made 
using available HST/ACS images which ultimately revealed their morphology as diffuse
and blended sources. The final list of our GC {\it candidates} contains 60 objects.

\begin{figure}
\begin{center}
\epsfig{figure=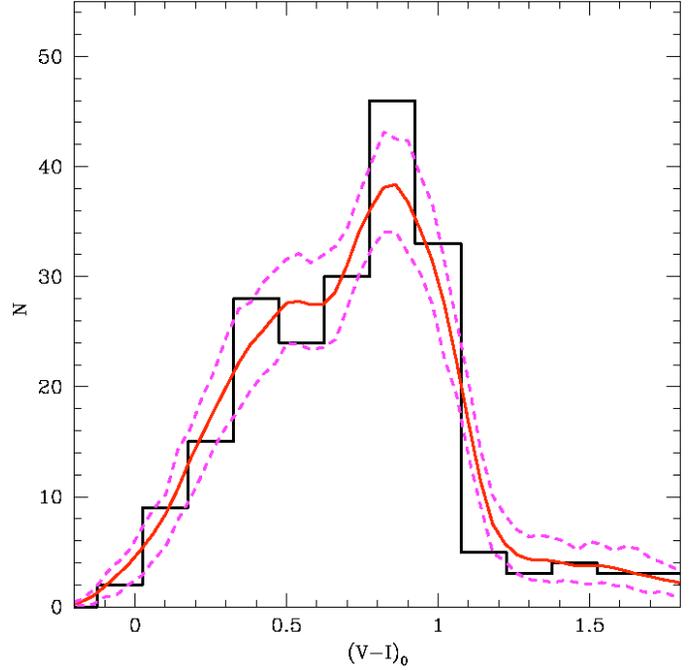, width=90mm}
\caption{\label{v-i_hist}$(V-I)_{0}$ color distribution of all point sources 
with $(U-B)_{0}\geq-0.3$ (see Fig.\,\ref{full}). The $U-B$ color cutoff was introduced 
to lower the contamination from faint or objects with highly uncertain photometry in
the starburst regions. The Epanechnikov-kernel probability density estimate and its 90\% confidence 
limits are plotted with thick/red and dashed/magenta lines, respectively. Similar color distributions are observed in star cluster systems of field dIrrs \cite{Sharina05}.}
\end{center}
\end{figure}

The $(V-I)$ color distribution for all point sources (presumably star clusters) 
with $(U-B)\geq-0.3$ is shown in Fig. \ref{v-i_hist}. The $U-B$ cutoff was introduced 
to lower the 
contamination from very faint unresolved amorphous objects with $V-I$ colors within 
the old GCCs range (compare with Fig.\,\ref{full}). An Epanechnikov-kernel probability 
density estimate shows a clear indication for (at least) two characteristic $(V-I)$ 
color peaks at $\sim\!0.4$ and 0.9 mag. These color peaks are similar to those found 
in the colors distributions of star cluster systems in field dIrr galaxies \cite[]{Sharina05}. 
However, the contamination effects in our sample prevent us from a more detailed analysis.

\subsection{Contamination and GC numbers}\label{fbc}
Despite the applied globular cluster selection criteria, there can still be 
sources contaminating our final GC candidate sample in the selected color
limits. Possibilities are faint ($V\geq21$\,mag) Galactic halo stars, 
distant and compact bulges of unresolved background galaxies, and unresolved 
low-redshift ($z\sim0.1-1.0$) starburst galaxies with ages $\leq300$\,Myr 
\cite[]{Puzia04}. Highly reddened ($A_{V}\geq1$\,mag) young compact sources 
(H\,{\sc ii} regions, OB associations, young star clusters) in the inner 
regions of NGC\,1427A are potential contaminants as well. This is estimated 
in Sect.\,\ref{fbc1} below.

Another source of contamination to the list of GC candidates of NGC\,1427A is 
contamination from possible Fornax intracluster globulars \cite[]{Bassino03} 
and the GC population of NGC\,1399 which is thought to extend to large distances 
\cite[]{Dirsch03}. We discuss this in Sect.\,\ref{fbc2}. 

\subsubsection{Foreground/background}\label{fbc1}

\begin{figure}[ht]
\epsfig{figure=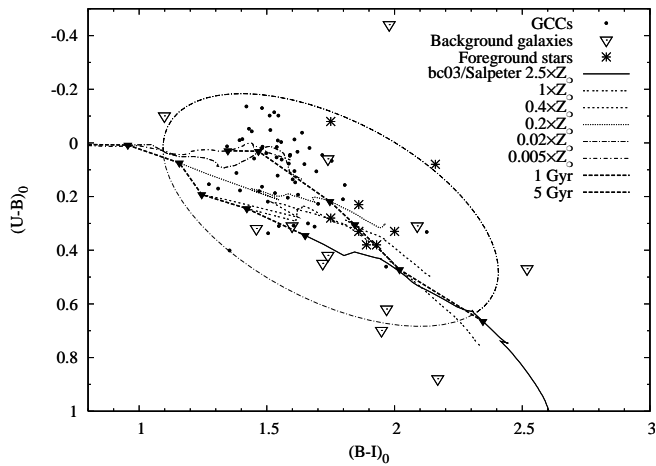, width=90mm}
\caption{\label{ubi}$(U-B)_0$\,vs.\,$(B-I)_0$ color-color diagram of the selected 
globular cluster candidates (dots), foreground stars (asterisk) and background 
galaxies from the FDF catalog (open triangles). The ellipse indicates the transformed 
GC candidate color-color selection region as indicated in Fig.\,\ref{full}. Lines 
show iso-metallicity tracks and isochrones for various ages and metallicities.}
\end{figure}

To estimate the foreground star contamination we used a synthetic Galaxy stellar 
population model\footnote{http://bison.obs-besancon.fr/modele/} \cite[]{Robin03}. 
For the specified color and magnitude ranges (Sect.\,\ref{cms} and \ref{pssc}) and 
within the FORS1 field of view the model predicts a total number of 7 foreground 
stars towards NGC\,1427A (see Fig.\,\ref{ubi}).

We estimate the contamination by background galaxies using the FORS Deep Field 
(FDF) data \cite[]{Heidt03}. The main criterion for selecting objects from the
FDF cataloged as galaxies is the limiting bright magnitude cutoff $I\geq20.1$\,mag 
and the completeness limit of our observations in the $I$-band (Fig.\,\ref{completeness}).  
However, the inspection of our final globular cluster candidate sample shows 
that there are no candidates with $I\geq24.5$\,mag, which actually represents the 
faint end of the GC luminosity function. Therefore, in Fig.\,\ref{ubi}, due to 
the lack of $V$-band photometry in the FDF, we present the $(U-B)_0$\,vs.\,$(B-I)_0$ 
color distribution of the background galaxies in the range $20.1\leq I\leq24.5$\,mag 
(triangles). It can be seen that $\sim7$ background galaxies are expected in the 
GC candidate color-color region.

\begin{figure}[ht]
\epsfig{figure=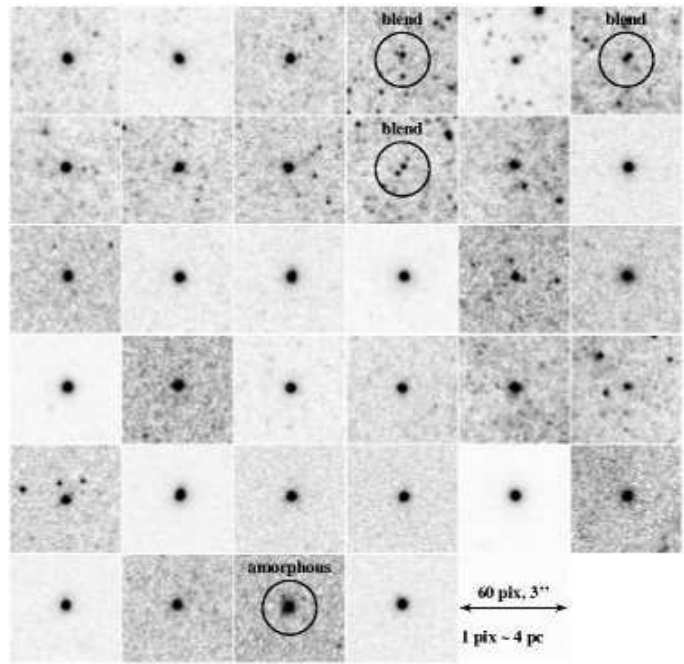, width=90mm}
\caption{\label{acs} F625W-band mosaic of the GC candidates as seen using HST/ACS. 
Circles indicate the three blends and the amorphous objects which were rejected 
from the original ground-based GC candidate list based on these ACS data.} 
\end{figure}

As an attempt to identify these background galaxies, as well as contamination from 
blended sources in the crowded galaxy regions, we take advantage of the high spatial 
resolution of archival HST/ACS images of NGC\,1427A taken with the F625W filter (Program 
GO-9689, PI: M. Gregg), which were combined using MultiDrizzle \cite[]{Koekemoer02}. 
A visual examination shows that none of the 34 GC candidates (out of 60) in common 
for the FORS1 and ACS fields are resolved as obvious background galaxies. The appearance 
of the GC candidates on the ACS chips are presented in Fig.\,\ref{acs}. This shows 
that our point-source selection criteria performed well and only a fraction of the 
seven BG galaxies estimated above could remain unresolved by ACS. However, three of 
our GCCs turned out to be blended sources and one seems amorphous (possibly a compact 
OB-association, see Fig.\,\ref{acs}). The latter is located in the inner starburst  
regions of the galaxy and passed the point-source selection criteria at the faint 
magnitude end. The blends and the amorphous object were excluded from subsequent 
analysis. 

Due to the smaller field of view of HST/ACS relative to the FORS1 field, not all 
candidates could be examined this way, especially those far from the galaxy body. 
Thus we can not extrapolate the above conclusions to the rest of the candidates, 
and unresolved background objects could still be present among the remaining 26 
GCCs. Discrimination between GCs and foreground stars is also not an easy task 
since one ACS pixel corresponds to $\sim$4\,pc projected size at the Fornax distance. 
Thus, GCs more compact than this would have a stellar appearance. Only spectroscopy 
can fully address the true foreground/background contamination fraction. A detailed 
analysis of the NGC\,1427A HST/ACS data will be presented in a forthcoming study.

\subsubsection{Fornax cluster GC contamination}\label{fbc2}
\begin{figure}[ht]
\epsfig{figure=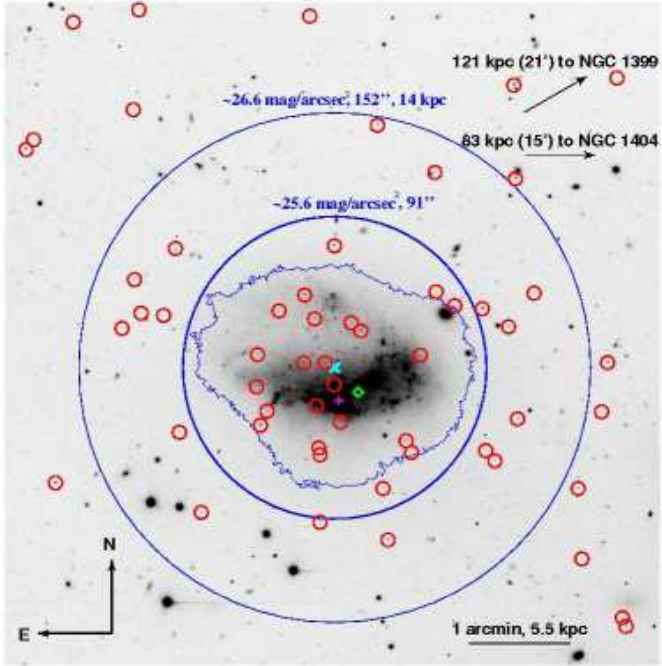, width=88mm}
\caption{\label{pos}$6\farcm8\times6\farcm8$ $V$-band VLT/FORS1 image. The positions 
of the selected (see Sect.\,\ref{pssc}\,and\,\ref{fbc1}) GC candidates are indicated
with (red) circles.  With arrows the directions and projected distances towards the 
Fornax cD galaxy NGC\,1399 and the elliptical NGC\,1404 are shown. With a ``$\times$'' 
the adopted center of NGC\,1427A is marked, with a ``+'' the fitted center derived from 
the median smoothed image and with a diamond the kinematical center determined from 
the ionized gas kinematics \cite[]{Chaname00}. The two large circles approximately 
encompass the $\mu_{V}$ isophotes at $\mu_{V}=25.6$\,mag/arcsec$^{2}$ (the true contours
of that isophote is shown as well) and at $\mu_{V}=26.6$\,mag/arcsec$^{2}$ 
(see Sect.\,\ref{sn} and Fig.\,\ref{SB}).}
\end{figure}

\begin{figure}[ht]
\epsfig{figure=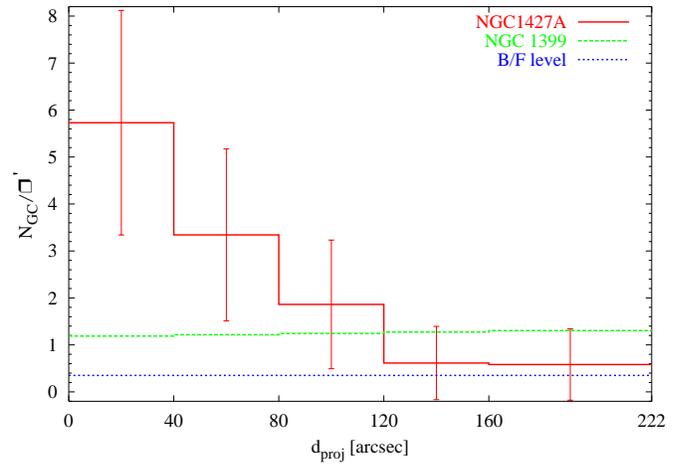, width=90mm}
\caption{\label{hist}Plot of the radial number density distribution of the NGC\,1427A 
GC candidates (thick line histogram) and the correspondent Poisson errors, the expected 
level of contamination by the NGC\,1399 GCS at the distance of NGC\,1427A (dashed/green 
line) and the background/foreground contamination level (dotted/blue line).} 
\end{figure}

Figure\,\ref{pos} shows the spatial distribution of GC candidates around NGC\,1427A
across the V-band FORS1 field of view. It can be seen that the number of GC candidates
increases towards the galaxy's center, which suggests that they are really associated 
with NGC\,1427A. The surface density of all GCCs is shown in Fig.\,\ref{hist} with 
red/solid histogram together with the Poisson errors.

However, the relatively close projected distances (see Fig.\,\ref{pos}) between
NGC\,1427A and the giant elliptical galaxy NGC\,1399 (22\farcm9) and the other 
nearby elliptical NGC\,1404 (15\farcm7) raise the question whether all found GC 
candidates really belong to NGC\,1427A. A study of the NGC\,1399 GCS by \cite{Dirsch03} 
concludes that even in their (most distant) studied field (which extends up to 
$\leq23$\,arcmin from NGC\,1399) the number counts have not yet reached the 
background value. However, as they point out, at such large distances their fields 
could contain possible intracluster GCs (ICGC) as proposed by \cite{Bassino03} in 
their search for GCs around dwarf elliptical galaxies in the Fornax cluster. The 
latter study derived an intracluster GC surface density of $\sim0.25$ and 0.13\,GCs/arcmin$^{-2}$ 
at distances of 40\arcmin and 110\arcmin from the cluster center, respectively. 
The existence of intracluster globulars is further supported by recent simulations 
by \cite{Yahagi05}.

The estimated GC number density contamination from the NGC\,1399 GCS is shown as 
a green/dashed histogram in Fig.\,\ref{hist}. At the distance of NGC\,1427A, 22\farcm9, 
$\sim1.26$ globular clusters per arcmin$^{2}$ are predicted \cite[]{Dirsch03}. 
Fig.\,\ref{hist} shows that at large galactocentric distance from NGC\,1427A, our 
number counts are considerably below the expected contamination from NGC\,1399, though 
still within the errorbars.

There are three possibilities that could account for this discrepancy between the 
observed and expected contamination level from NGC\,1399. The first one is that we 
are missing a number of GCs. However, according to the completeness analysis 
(Sect.\,\ref{compl}) we are confident that this is not the case. The second possibility 
is that the \cite{Dirsch03} values might not be aplicable toward NGC\,1427A, which 
is likely, because the fields studied by \cite{Dirsch03} extend towards the nearby 
elliptical galaxy NGC\,1387 (their Fig.\,1) which could give rise to an enhanced 
GC number density. The third possibility is to suggest a non-spherical distribution 
of GCs around NGC 1399 or ICGCs, resulting in a lower number density towards NGC 
1427A as compared to other directions. We think that a combination of the last two 
is the more plausible explanation. A possible GC contamination from the even closer 
(15\farcm7) elliptical NGC\,1404 seems to be ruled out as well, since its GCS is not 
as extended as that of NGC\,1399. Indeed, at a distance from NGC\,1404 of only 3\,arcmin 
the number density is already 0.63\,arcmin$^{-2}$ \cite[]{Richtler92,Forbes98} which, 
when extrapolated to the position of NGC\,1427A, becomes irrelevant.

The estimated total number density of background/foreground objects (Sect.\,\ref{fbc1}) 
shown in Fig.\,\ref{hist} ($0.35\pm0.59$\,objects/arcmin$^{2}$) is below the levels 
we reach in our outermost radial bins ($0.61\pm0.78$\,GCCs/arcmin$^{2}$). Although 
this values are within the errorbars the difference could be interpreted as: ($i$) 
either the local BG/FG is enhanced, or ($ii$) that there exist genuine GCs associated 
with either NGC\,1427A, NGC\,1399 or the ICGSs population at those galactocentric distances. 
Although the latter possibility seems more plausible to us, this is irrelevant for 
our purposes because the maximum contamination that we could possibly assume would be 
what we measure at the outermost bins (ideally containing no GCs from NGC\,1427A). 
Thus we adopt $0.61\pm0.78$\,counts/arcmin$^{2}$ as our best-estimate for the contamination 
level.

\subsubsection{Final GCC number}\label{fbc3}
Recalling the total of 60 candidates (Sect.\,\ref{pssc}) here we determine what 
fraction of these are contaminating objects. As was shown in Sect.\,\ref{fbc1} we 
expect 7 foreground stars, which are indistinguishable from actual GCs with our 
data. The expected number of background galaxy contamination (Sect.\,\ref{fbc2}) 
is a fraction of the 7 objects predicted by FDF. Therefore we end up with a number 
between 10 and 14 BG/FG contaminants. 

In Sect.\,\ref{fbc3} we estimated a maximum density of $0.61$\,objects/arcmin$^{2}$ 
contamination, which when multiplied by the 43\,arcmin$^{2}$ FORS1 field of view, 
results in 26 contaminating objects. Therefore we expect a total number of contaminants 
between 10 and 26. Taking the mean value of 18 and the four objects rejected by ACS 
(Sect.\,\ref{fbc1}) the final number of GCCs is $N_{GC}=60-4-18=38\pm8$, where 
the error encompasses the minimum and maximum number of contaminants quoted above.

\section{Analysis of the GC candidates' properties}\label{analysis}

\subsection{Colors}\label{colors}
In a previous study of NGC\,1427A with the 2.5m DuPont telescope at Las Campanas 
Observatory by \cite{Hilker97} a selection of globular cluster candidates on the 
basis of their $V-I$ colors only was made. The introduction of the $U-B$ color in 
our study significantly improved the discrimination between globular clusters and 
highly reddened young star clusters. Only 7 globulars are in common between the 
current and the previous study. The larger number of GC candidates we derive is 
due to the significantly larger field of view of the FORS1 CCD detector and the 
much deeper exposures.

\begin{figure}[ht]
\begin{center}
\epsfig{figure=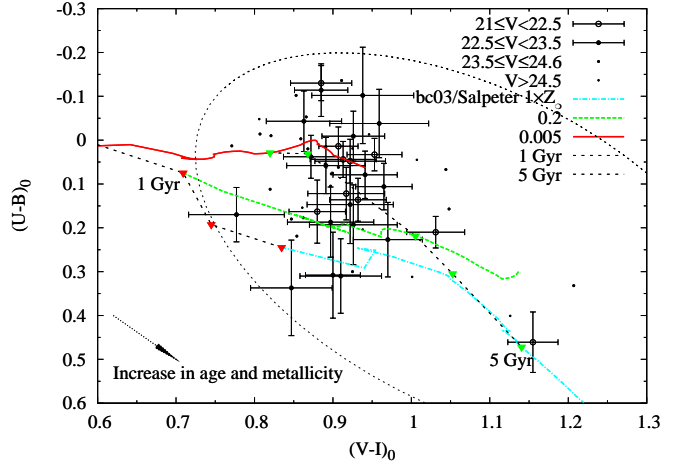, width=90mm}
\caption{\label{ubvi}The final selected globular cluster candidate sample.
Dashed lines connect the SSP models with different metallicities (in solar units) 
for ages 1\,Gyr and 5\,Gyr. With an ellipse is shown the GC color selection locus.}
\end{center}
\end{figure}

The final colors of the GC candidates are shown in Fig.\,\ref{ubvi}. They 
cluster around low metallicities ($Z=0.0001$ and 0.0004) and old or intermediate 
ages ($\sim5$\,Gyr) according to the \cite{BC03} SSP models. With dashed lines 
the 1 and 5\,Gyr isochrones connecting SSP models with different metallicities 
are indicated. The photometric errors at these magnitudes and the convergence 
of the theoretical models prohibit accurate age and metallicity derivations for 
the individual globular cluster candidates. However, we can conclude that the 
ages and metallicities of the GC candidates around NGC\,1427A as whole are 
comparable to the ages and metallicities of the old metal-poor Milky Way GCs.

The mean color $(V-I)_0=0.92\pm0.08$ of all GC candidates is similar to what was 
found for the globular cluster systems in dIrr, dSph and dE galaxies \cite[]{Sharina05, 
Seth04, Lotz04}. This color corresponds to the blue GC population found in giant 
early-type galaxies and also obeys the relation between the host galaxy luminosity 
and the mean GC color (NGC\,1427A: $M_{V}=-18.13$, see Sect.\,\ref{sn}) \cite[]{Harris05, 
Peng05, Larsen01, Lotz04}.

\begin{figure}[ht]
\epsfig{figure=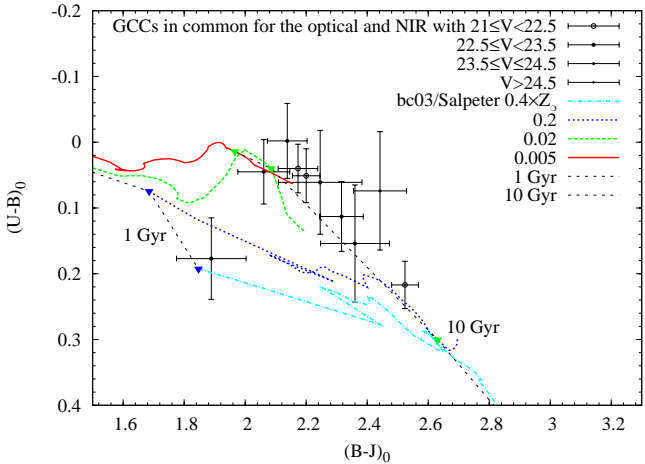, width=90mm}
\caption{\label{IR}Near IR color-color diagram for GC candidates in common for 
the FORS1 and ISAAC fields. Overplotted are the \cite{BC03} SSP models for ages 
$\geq1$\,Gyr and metallicities lower than $0.4\times Z_{\odot}$. With dotted 
lines the 1 and 10\,Gyr isochrones are indicated.}
\end{figure}

In Figure\,\ref{IR} we show the properties of the GC candidates in 
the $(U-B)_0$\,vs.$(B-J)_0$ color-color plane. This color-color selection provides 
the best age/metallicity discrimination governed by the filters' transmission 
curves. The different line types in Figure\,\ref{IR} represent \cite{BC03} SSP models 
formetallicities lower than $0.4\times Z_{\odot}$ ($Z=0.008$). Although the IR images 
were not deep enough, a simple comparison with the 1 and 10\,Gyr isochrones (dotted 
lines in Fig.\,\ref{IR}) reveals the predominantly old ages ($\geq10$\,Gyr) and low 
metallicities ($\leq0.2\times Z_{\odot}$) of the GC candidates. \cite{Hilker97} showed 
that in integral properties NGC\,1427A is similar to the Large Magellanic Cloud (LMC). 
Hence, taking into account the LMC metallicity of $Z=0.04\times Z_{\odot}$ we can set 
the expected upper metallicity limit to this value. This is consistent with what is 
observed in Fig.\,\ref{IR}.

\subsection{Luminosity function}\label{LF}

\begin{figure}[ht]
\epsfig{figure=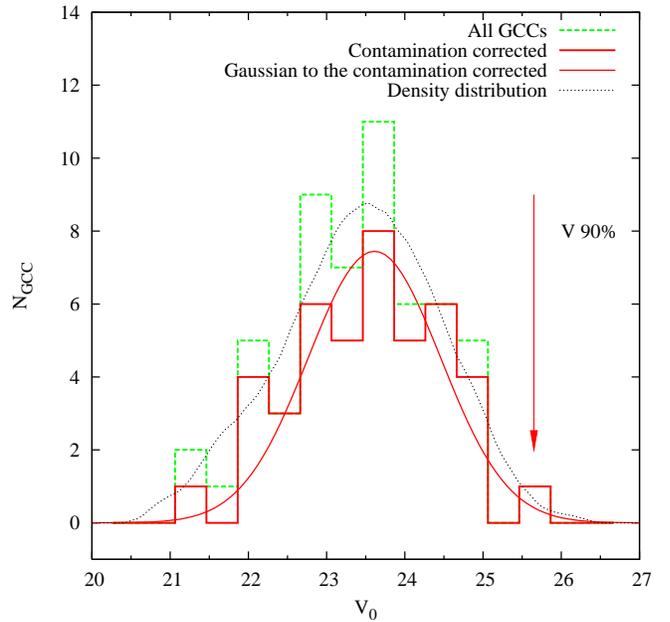, width=120mm}
\caption{\label{gclf}NGC\,1427A globular cluster luminosity function (GCLF). The solid 
(red) histogram represents the contamination corrected GCLF, the dashed histogram the 
uncorrected one. The curves are a Gaussian fit to the corrected GCLF (solid line) and 
the probability density estimate of the uncorrected GCLF using an Epanechnikov kernel 
(dotted curve). The arrow indicates the 90\% completeness limit in V-band.}
\end{figure}

Figure\,\ref{gclf} shows the NGC\,1427A globular cluster luminosity function (GCLF) 
of the original and contamination corrected GCCs sample in bins of 0.4\,mag. The 
dashed (green) histogram shows the GCLF for the original GC candidate sample while 
the solid (red) histogram represents the number counts corrected for contamination 
using the luminosity distribution of the contaminating objects in the outermost bins 
as estimated in Sect.\,\ref{fbc2}.

Globular cluster luminosity functions have been studied for many galaxies \cite[eg.][and 
ref's therein]{Ashman&Zepf98, Kissler-Patig00, Richtler03, McLaughlin03}.
It seems that the shape of the GCLF is universal among ``old'' GC systems and can
be represented by a Gaussian or a $t_5$ function. In Fig.\,\ref{gclf} the
Gaussian fit to the corrected GCLF is shown. Due to the low number statistics we kept 
the width of the Gaussian fixed to $\sigma=1.2$ which is a typical value that fits 
most of the studied galaxies' GCLFs \cite[eg.][]{Barmby01}. This choice of the dispersion 
gave $V_{TO}=23.61\pm0.10$ for the turnover magnitude of the BG/FG corrected sample. 
If we do not fix $\sigma$ we obtain $\sigma=1.45\pm0.16$ and $V_{TO}=23.59\pm0.11$. 
The probability density function of the non-parametric density estimate using an 
Epanechnikov kernel (dotted line in Fig.\,\ref{gclf}) of the uncorrected GCLF has 
its maximum value at $V_{TO}=23.53\pm0.07$\,mag. The quoted uncertainties above 
include only the photometric errors and the uncertainty in the fit. The effect of 
the uncertainty in the number counts is reflected in the small difference in the 
final turnover magnitudes of the two distributions, which is within the errorbars.

Due to its universality the GCLF was frequently used as a distance indicator 
\cite[for details see][]{Richtler03}, originally proposed by \cite{Hanes77}. Assuming 
a universal turnover magnitude of $M_{V}=-7.40\pm0.11$, as derived for the MW GCS 
\cite[]{Harris01} and those of other galaxies \cite[eg.][and ref's therein]{Larsen01, 
Ashman&Zepf98}, the distance modulus to NGC\,1427A is $(m-M)=31.01\pm0.21$\,mag 
($15.9\pm1.6$\,Mpc). Our most shallow observation is in $U$-band with the 90\,\% 
completeness limit at $U=26.21$, which is however still $\sim1.4$\,mag deeper than 
the expected $U_{TO}\sim24.8$\,mag (assuming a mean $U-B\sim0.1$, $(B-V)\simeq0.7$, 
$M_{V_{TO}}=-7.40$, and a distance modulus of 31.39\,mag). The last bin of the histogram 
($V\simeq25.8$\,mag) corresponds to $U\simeq26.6$\,mag which is approximately at the 
85\% completeness level. This implies a correction which makes the $V_{TO}$ fainter 
by 0.01\,mag, which is well within the error and therefore we do not consider it further.

A recent study by \cite{DiCriscienzo05} presents a new calibration of the GCLF 
performed on the basis of RR Lyrae in the MW, M31 and close-by galaxies as ``primary'' 
standard candles. They derive a GCLF turnover magnitudes of $V_{TO}=-7.66\pm0.11$ 
for the MW, $V_{TO}=-7.65\pm0.19$ for M31 and $V_{TO}=-7.67\pm0.23$ for the other 
nearby galaxies. The weighted mean for the combined turnover magnitude then is 
$M_{V_{TO}}=-7.66\pm0.09$ mag. If we apply this value to the GCS of NGC\,1427A we 
obtain a distance modulus of $(m-M)=31.25\pm0.20$\,mag ($17.8\pm1.7$\,Mpc).

Finally, independently of which absolute GCLF turnover magnitude is adopted, the 
direct comparison of the apparent GCLF turnovers 
for NGC\,1399 \cite[$V_{TO}=24.01\pm0.1$,][]{Dirsch03} and NGC\,1427A places 
the latter approximately at $3.2\pm2.5\,({\rm statistic})\pm1.6\,({\rm systematic})$\,Mpc 
in front of NGC\,1399. The statistical error is determined from the photometric 
and our fitting routine uncertainities of the turnover magnitude; the systematic 
error includes the uncertainties of the absolute magnitude and the GCLF method 
itself \cite[]{Richtler03}. It should be noted that the giant elliptical NGC\,1404 
is located even closer in projection to NGC\,1427A. Unfortunately, the GCLF turnover 
magnitude for NGC\,1404 \cite[]{Blakeslee&Tonry96, Richtler92} is not as accurately 
determined as that of NGC\,1399, and thus we can not perform a similar comparison.

\subsection{Specific frequency}\label{sn}

The 'specific frequency' ($S_{N}$) is a quantity introduced by 
\cite{Harris&vandenBergh81} to intercompare the GCSs between different 
elliptical galaxies. The $S_{N}$ is defined as the total GC ($N_{\rm{GC}}$) 
population normalized to a galaxy luminosity of $M_{V}=-15$\,mag:

\begin{equation}
S_N = N_{\rm{GC}}10^{0.4(M_{V}+15)}
\end{equation}

In order to derive the specific frequency properly one needs to correct for 
background and foreground contamination, the unobserved part of the GCS (and
the parent galaxy) and the completeness of the observations at the faint end 
of the GCLF. One of the most important issues is that $S_{N}$ should be derived 
from GC counts and total galaxy light both estimated within the same area. 
Otherwise systematics in the GCS and/or galaxy profile could affect the estimated 
$S_{N}$ values \cite[]{Dirsch03, Ostrov98}.

\begin{figure}[ht]
\epsfig{figure=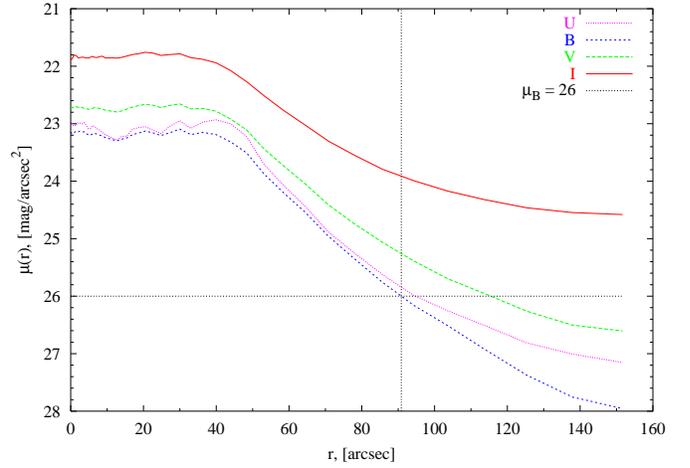, width=90mm}
\caption{\label{SB} The NGC\,1427A surface brightness profile in $U,B,V,{\rm and} 
I$-bands. The vertical dotted line indicates the radius of the $\mu_{B}=26$\,mag/arcsec$^{2}$ 
isophote, at which the $V$-band $\mu_{V}=25.6$\,mag arcsec$^{-2}$ was determined 
and plotted in Fig.\,\ref{pos}.} 
\end{figure}

Having the $V$-band 90\,\% completeness limit $\sim1.5$\,mag (see Sect.\,\ref{LF} 
and Fig.\,\ref{completeness}) beyond the expected GCLF turnover magnitude, 
completeness corrections are insignificant for the estimate of $S_{N}$. Due 
to the complete coverage of the galaxy including its outer parts we do not need 
to make any geometrical corrections. In order to measure the total galaxy 
light of NGC\,1427A we used the IRAF/STSDAS task ELLIPSE on the background 
subtracted images. The bright, obvious background and foreground sources were 
masked out. Due to the complex morphology of this irregular galaxy the definition 
of its center was a complicated task. We adopted the center which best represents 
the diffuse galaxy light distribution at the radius of the $\mu_{B}=26$\,mag 
arcsec$^{-2}$ isophote (marked with ``$\times$'' in Fig.\,\ref{pos}). This is 
the common center for the both circles drawn in Fig.\,\ref{pos} in which we 
estimate the $S_{N}$. The V-band isophote ($\mu_{V}=25.6$\,mag arcsec$^{-2}$ ) 
shown in Fig.\,\ref{pos} is determined at the radius of the B-band $\mu_{B}=26$ 
isophote (compare with Fig.\,\ref{SB}). Another approach was to fit the center 
on the 41\,pix ring aperture smoothed image (see Sect.\,\ref{redphot}) using the 
IRAF centroid algorithm. The result of this centering is shown with a ``+'' symbol 
in Fig.\,\ref{pos}. Based on the NGC\,1427A ionized gas kinematics, \cite{Chaname00} 
defined a kinematical center which is indicated with a diamond symbol in Fig.\,\ref{pos}. 
The latter two center coordinates are offset with respect to the center defined 
by the diffuse galaxy light while the adopted geometrical V-band isophote center 
does not only represent the center of the diffuse galaxy light (stars) but also 
is the most symmetrical center of the GC candidate distribution around NGC\,1427A. 
In order to check the centering effect we measured the differences of the total 
galaxy magnitudes assuming the different centers and found negligible deviations 
below 2\,\% (0.02\,mag). Thus we adopted the isophotal center as a good enough 
approximation.  
In Fig.\,\ref{SB} the galaxy surface brightness profiles for $U,B,V$, and $I$ are 
shown. With the dotted lines the extension of the $\mu_{B}=26$ isophote (at $r_{1}=90\farcs9$) 
is indicated, for which the total magnitude $V_{TOT,1}=13.06\pm0.02$\,mag was 
determined. As can be seen, the surface brightness profile reaches the background 
level in the I-band approximately at $r_{2}=151\farcs7$. We adopt this truncation 
radius for the other pass-bands. The total magnitude at that radius 
would be $V_{TOT,2}=12.88\pm0.02$\,mag. The determination of $S_{N}$ for both 
radii will give us information on the radial dependence of the specific frequency 
in NGC\,1427A. Inside $r_{1}$ and $r_{2}$ we count 23 and 40, respectively.

The correction for contamination is based on the determined level of 
0.61\,objects/arcsec$^{-2}$ (see Sect.\,\ref{fbc2}). Thus, for the two radii 
we determine a correction of 4 and 12 contaminants leading to $N_{\rm{GC}}=19$ 
and 28, respectively. 

Using the adopted distance to Fornax we obtain $M_{V,1}=-18.33\pm0.20$ and $M_{V,2}=-18.51\pm0.20$\,mag. 
The calculated specific frequencies are $S_{N}=0.88$ and $1.09\pm0.23$. Using the 
distance modulus determined from the GCLF turnover we compute $M_{V,1}=-17.95\pm0.21$ 
and $M_{V,2}=-18.13\pm0.21$\,mag, which results in the following specific frequencies, 
$S_{N}=1.3$ and $1.6\pm0.23$ for the smaller and larger aperture, respectively. These 
$S_{N}$-values should be considered as lower limits for NGC\,1427A's specific frequency 
since an age-fading of the galaxy light should be applied to compare it with the 
$S_{N}$ of giant early-type galaxies. As \cite{Miller98} estimated, the final total 
$V$ magnitude of a galaxy that is forming stars for 5\,Gyr at constant rate and then 
is fading for another 5\,Gyr will be fainter by $\sim1.5$\,mag as during the star 
forming phase. Assuming that the number of old GCs is conserved, this luminosity 
evolution will increase the $S_{N}$ by a factor of $\sim4$ during this time. Also, 
the number of the old GCs could increase of the most massive and/or compact young 
star clusters, present in the starburst regions, survive the disruptive dynamical 
evolution.

The $S_{N}$ value measured here ($\geq1$) is higher than typical values for dIrrs 
\cite[e.g. $S_{N}\simeq0.5$ for LMC,][]{Harris91} in the Local Group. However, our 
estimates are comparable with results by \cite{Seth04} for dIrrs in the Virgo and 
Fornax clusters, $S_{N}>2$, which suggests that the environment influences the GC 
formation efficiency in dIrr galaxies.

Another approach avoiding the uncertain galaxy light age-fading estimate is to 
use the $T$ parameter proposed by \cite{Zepf&Ashman93} which relates the
total GC number to the total galaxy mass:
\begin{equation}
T=\frac{N_{\rm{GC}}}{M_{G}/10^{9}M_{\odot}}
\end{equation}
where $M_{G}$ is the total galaxy mass. \cite{Chaname00}, using a rigid-body rotation 
model, determined the NGC\,1427A angular velocity of $\omega=12.8\pm1$\,km\,$\rm s^{-1}$\,$\rm kpc^{-1}$. 
They derive a lower limit of the dynamical mass of $M_{G}=(9\pm3)\times10^{9}M_{\odot}$ 
within 6.2\,kpc radius corresponding to $\mu_{V}=24.7$\,mag/arcsec$^{2}$. This isophote 
corresponds to 77\farcs8 distance from the dynamical center. Within this aperture, 
after correction for the contamination of 3 objects, we count 19 GC candidates and 
thus obtain the local $T$ value of $2.1\pm1$. However, the total T value should be 
used for comparison. We calculate at $r_{2}=151\farcm7\simeq12$\,kpc the NGC\,1427A 
dynamical mass of $M_{G}=(6\pm3)\times10^{10}M_{\odot}$ and $N_{\rm GC}=28$, which 
gives $T=0.5\pm1$. This GC formation efficiency per unit mass is smaller but comparable 
to the $T$ values of spiral galaxies \cite[e.g.,][]{Chandar04, Goudfrooij03, Barmby&Huchra01, 
Kissler-Patig99b}.

\section{Conclusions}\label{conclusions}
Based on deep VLT observations, we study the old GC population of the dwarf irregular 
galaxy NGC\,1427A. 
The results of our analysis are as follows. After applying various selection 
criteria, namely color cuts, point source selection, exclusion of H${\alpha}$-emitting 
sources, and a visual inspection of HST/ACS images resulted in 56 GCCs. Accounting 
for contamination (Sect.\,\ref{fbc}) this finally leads to $38\pm8$ globular cluster 
candidates, likely associated with NGC\,1427A. Their radial number density distribution 
is concentrated towards the galaxy center, which suggests that they belong to this 
galaxy. The colors of the selected GC candidates show that most of them are metal-poor 
($Z\leq0.08\times Z_{\odot}$), old globular clusters ($\geq5$\,Gyr) according to comparisons with 
simple stellar population models \cite[]{BC03} and colors of the old GCs in the 
Milky Way. This is in agreement with results from previous investigations on GCS 
in dIrr, dSph and dE galaxies \cite[e.g.][]{Sharina05, Seth04, Lotz04}, which show 
that the GCSs in these systems share similar properties, hence similar early star 
formation histories.

Using the GCLF turnover magnitude as a standard candle, we estimated the distance 
to NGC\,1427A for the first time. We derive a distance modulus of $(m-M)=31.01\pm0.21$\,mag 
($15.9\pm1.6$\,Mpc) taking the measured GCLF turnover of $V_{TO}=23.61\pm0.11$ and assuming 
$M_{V_{TO}}=-7.40\pm0.11$ as universal \cite[]{Harris01}. If we use the RR\,Lyrae calibrated 
GCLF $M_{V_{TO}}=-7.66\pm0.09$ \cite[]{DiCriscienzo05} we obtain $(m-M)=31.25\pm0.20$ mag 
($17.8\pm1.5$\,Mpc). The analysis shows that NGC\,1427A is placed $3.2\pm2.5\,({\rm statistic})\pm1.6\,({\rm systematic})$\,Mpc 
in front of the giant cD galaxy NGC\,1399. If NGC\,1427A really is located more than 
2-3 Mpc away from the cluster center, the suggested interaction of NGC\,1427A with the 
dense intracluster medium triggering the intense star formation activity would then 
probably occur at large cluster-centric radii. This result, however, should be taken 
with precaution considering the uncertainties involved in the GCLF turnover point and 
the possible systematics in the method itself of about 0.2\,mag \cite[for a review see][]{Richtler03}. 
More importantly, a precise analysis of the relationship between NGC\,1427A and the 
cluster environment must involve the giant elliptical NGC\,1404 as well. We are currently 
prevented to do this due to the large uncertainties in the relevant distances. However, 
our relative distance result supports the \cite{Drinkwater01} finding that the kinematically 
distinct population of infalling dwarf galaxies in Fornax shows an extended spatial 
distribution.

We obtained a present-day specific frequency of $S_{N}=1.6\pm0.23$. However, this
galaxy still is actively forming stars. Hence, an age fading of the galaxy's
light should be applied prior to compare its $S_{N}$ value to those of `old'
early-type galaxies. Applying such a correction \cite[according to][]{Miller98} 
would lead to a $S_{N}$ value of $\sim 7$ after passive evolution for a few Gyr.
The large $S_{N}$ value suggests that dIrrs could contribute to the blue GC population 
and its total numbers in giant early-type galaxies through dissipationless merging or 
accretion. Since they are still actively forming stars and star clusters, they may also 
contribute to the red GC population by some newly-formed GCs. Those clusters might 
form out of material that was enriched during the starburst event, probably triggered 
by the interaction with the cluster environment. 

Given the current paucity of deep imaging studies of dIrr galaxies combined  
with the fact that they are likely building blocks of giant galaxies in the
context of the hierarchical merging scenario, we suggest that detailed studies
of dIrr galaxies in a range of environments be undertaken to establish their 
range of GCS properties as a function of environment density. This should
provide important constraints on the formation of these galaxies 
as well as on their global impact on galaxy formation and evolution.

\acknowledgements
IG acknowledges the partial support from the DFG Graduate Research School 787 on 
``Galaxy Groups as Laboratories for Baryonic and Dark Matter''. IG and PG would 
like to thank the director of STScI for the award of a graduate studentship funded 
by the Director's Discretionary Research Fund. The authors also gratefully acknowledge 
the valuable comments and discussions with Rupali Chandar, Michael Drinkwater and 
Klaas de Boer. JC wishes to express his gratitude to the faculty at the Department 
of Astronomy of The Ohio State University for their continuous support to this external 
collaboration. AR and LF would like to acknowledge the support from FONDECYT Regular 
Grant 1020840 and the FONDAP center for Astrophysics.
\bibliographystyle{aa}
\bibliography{citations}

\end{document}